\documentstyle[12pt,iopfts]{ioplppt}
\eqnobysec
\begin{document}
\title{Quantization of massless fields \\
over the static Robertson-Walker space \\
of constant negative curvature}[Quantization of massless fields]
\author{S A Pol'shin\ftnote{1}{E-mail: itl593@online.kharkov.ua}}
\address{Institute for Theoretical Physics \\
NSC Kharkov Institute of Physics and Technology \\
Akademicheskaia St. 1, 61108 Kharkov, Ukraine }
\jl{6}
\date{}

\begin{abstract}
Taking the ${\Bbb R}^1 \times H^3$ space  as 
an example, we develop the new method of quantization of fields 
over symmetric spaces.
We construct the quantized massless fields of an arbitrary spin over the ${\Bbb R}^1 \times H^3$
 space by the resolution over the systems of "plane waves" which are
solutions of the corresponding wave equations. The propagators of these
fields are ${\Bbb R}^1 \times SO(3,1)$-invariant and causal. For spin 0 and
1/2 fields the propagators are obtained in the explicit form.
\end{abstract}

\section{Introduction}

As the Minkowski space is only an idealization of a real curved space, then 
the problem of construction of QFT over the curved space appears naturally.  
The solution of this problem shall be most  complete and formally similar to
that over the flat space if we restrict ourselves to the symmetric spaces
with  symmetry groups of a sufficiently high  range. To this end in~\cite{dS-JPA} we applied the method
of generalized coherent states (CS) to the quantization of massive spin 0 and 1/2 fields over de Sitter 
space. Another  curved space in which  we can apply this method 
 is the ${\Bbb R}^1 \times H^3$ space. Like de Sitter space, 
it is the particular case of Robertson-Walker spaces of  constant curvature; from the other hand,
it belongs to the class of
so-called ultrastatic spaces which is also important as an arena of QFT~\cite{Byts}.
 In~\cite{MPLA} 
using the CS method we found the solutions of wave equations for the massless 
particles of arbitrary spin over the ${\Bbb R}^1 \times H^3$ space in the 
form of so-called "plane waves", and using these "plane waves" for the spin 
zero particles we proposed a new method of quantization of spin zero massless 
field over this space.  In the present paper we apply the CS 
method to the quantization of massless fields of arbitrary nonzero spin over this 
space. The previous attempts to construct  the spinor propagators over the 
${\Bbb R}^1 \times S^N$~\cite{Inagaki} and $H^N$~\cite{Muck} spaces 
concerned with the massive spin 1/2 fields only and did not yield the results 
of a clear physical and mathematical sense. General results on the 
quantization of massive spinor and vector fields over static space-times may be found 
in~\cite{jin}.

The present paper is constructed as follows. In section 2 we state the well
known results concerning the Lorentz group and  its
irreducible representations. In section 3 we consider the scalar CS system
for the Lorentz group which is a generalization of that considered
in~\cite{coher1} onto the arbitrary values of parameter. In section 4 we
consider the spinor CS system for the Lorentz group introduced implicitly by
S.Weinberg in~\cite{48}. Following~\cite{MPLA}, we construct the system of
solutions of wave equations for the massless particles of arbitrary spin
over the ${\Bbb R}^1 \times H^3$ space as the product of this CS system and
that constructed in section 3, and find the invariance properties of this
system of solutions. In section 5 we construct the quantized massless fields
of arbitrary spin  by the resolution over the system of solutions constructed
in section 4, and show that these fields have the
 ${\Bbb R}^1 \otimes SO(3,1)$-invariant propagators, and find
the explicit form of these propagators in   the case of spin 0 and 1/2 fields. 
In Section 6 we compare our results with other approaches to the quantization of
fields in curved spaces.
In Appendix A the arbitrary spin two-point functions over $H^3$ space
are derived. In Appendix B  the causality of propagators  for an
arbitrary spin constructed in Section 5 and analyticity of corresponding
two-point functions  are proved.  In Appendix C we show that these propagators 
possess the correct Minkowski space limit. 

\section{The $H^3$ space and its symmetry group}

The three-dimensional hyperbolic space
$H^3$ is the hyperboloid of the  radius $R$ in a
 four-dimensional Lorentzian space with the metric tensor
$\eta_{\alpha\beta}=
\mbox{diag }(+1,+1,+1,-1),\ \alpha,\beta=1,\ldots,4$,
determined by the equation
$\eta_{\alpha\beta}x^{\alpha}x^{\beta}=-R^{2}$.
We denote $\kappa_{\bi x}=\sqrt{1+{\bi x}^{2}/R^{2}}.$
The symmetry group of the $H^3$ space is $SO(3,1)$ with the
generators $J_{\mu\nu}=-J_{\nu\mu}$ 
and its commutation relations are
\[{[}J_{\mu \nu}, J_{\rho \sigma}{]}=\eta _{\mu \sigma}J_{\nu \rho}+
\eta _{\nu \rho}J_{\mu \sigma}-\eta _{\mu \rho}J_{\nu \sigma}-
\eta _{\nu \sigma}J_{\mu \rho}.\]
The irreducible representations
 of the Lorentz group are well-known~\cite{49}.
They are determined by  two complex numbers $j_+$ and $j_-$ 
obeying the condition
\[ j_{+}-j_{-}=\pm s, \quad s=0, 1/2, 1,\ldots \]
where $s$ is a spin. For the case when $j_+$ and $j_{-}$ are integer or
half-integer   the corresponding representation is finite-dimensional and
non-unitary. In the opposite case it is infinite-dimensional and unitary.

To each element $g\in SO(3,1)$ we connect the matrix
$\left(
\begin{array}{ll}
a & b \\ c & d
\end{array}
\right)$
belonging to the $SL(2,{\Bbb C})\simeq SO(3,1)$ group. Then we can define the
action of this group over $\Bbb C$:
\[z \mapsto z_{g}=\frac{az+b}{cz+d}.\]
Then the representation $(j_+ ,j_-)$ admits a realization over the functions
on $z$ and $\bar z$~\cite{49}:
\begin{equation}\label{lor1}
T_{j_+ j_-}(g) f(z,\bar{z})=(cz+d)^{2j_+} (c^* \bar{z}+d^* )^{2j_-}
f(z_g ,\bar{z}_g).
\end{equation}
For the finite-dimensional irreducible representations the function
$f(z,\bar{z})$
should be a polynomial on $z$ and $\bar{z}$ of the power equal or less
than $2j_+$ and $2j_-$, respectively.  Hereafter we shall deal with the
finite-dimensional representations $(s,0)$; their generators are
\begin{equation}\label{pi-s}
J_{i0}=\frac{\i}{2}\varepsilon_{ikl}J_{kl}=S_i
\end{equation}
where $S_i$ are  generators of the rotation subgroup:
\begin{equation}\label{lor2}
S_+ =-\frac{\partial}{\partial z} \qquad
S_- =z^2 \frac{\partial}{\partial z} -2sz \qquad
S_3 =s -z\frac{\partial}{\partial z}.
\end{equation}
We denote as $L_{\alpha\beta}(g)$ the matrix of
orthogonal transformation corresponding to the element $g\in SO(3,1)$
and define the action of the Lorentz group over the $H^3$ space as
$x^{\alpha}\mapsto x^{\alpha}_{g}=L^{\alpha}_{\ \beta}(g)x^{\beta}$.
The stationary subgroup of an
arbitrary point of the $H^3$ space is $SO(3)$; then we can identify this
space with the coset space $SO(3,1)/SO(3)$.

\section{Scalar coherent states}

Let us consider an arbitrary light-like four-vector:
\begin{equation}\label{lor5}
 n^\alpha =(\omega{\bi q},\omega) \qquad {\bi q}^2 =1.
\end{equation}
Define the orthogonal action of the Lorentz group on it:
\[ n^\alpha \mapsto n^\alpha_g =L^\alpha_{\ \beta}(g)n^\beta \]
and  represent again the resulting vector in the form of~(\ref{lor5}).
 This yields the projective action of the Lorentz group over the unit
three-vector: ${\bi q}\mapsto {\bi q}_g$.

Let us consider the representation of the Lorentz group acting over the
functions depending on $\bi q$:
\begin{equation}\label{4. 30}
T_\sigma (g)f({\bi q})=
\left( \frac{n^4_g}{n^4}\right)^\sigma f({\bi q}_{g^{-1}})
\end{equation}
where $\sigma\in {\Bbb C}$. At
 $\sigma=\i\omega R-1$ the above representation is the
infinite-dimensional and irreducible one for the spin zero particles.
In~\cite{coher1} was shown that in this case the corresponding CS
system for the $H^3$ space is
\[ \phi_{\bi q}({\bi x};\sigma)=
\left( \kappa_{\bi x}-\frac{{\bi q}{\bi x}}{R} \right)^{\sigma}. \]
This result may be proved for an arbitrary $\sigma\in {\Bbb C}$ in the quite
similar way.
Then~(\ref{4. 30})  yields the following transformation property
of the constructed CS system for an arbitrary $\sigma$:
\begin{equation}\label{transf-phi}
\phi_{\bi q}({\bi x}_g;\sigma)=\left( \frac{n^4_g}{n^4}\right)^\sigma
\phi_{{\bi q}'}({\bi x};\sigma) \qquad {\bi q}'={\bi q}_{g^{-1}}.
\end{equation}
It is easily seen~\cite{MPLA} that the functions
\begin{equation}\label{psi-0}
 \psi ({\bi x},t)=\e^{\i\omega t}\phi_{\bi q}({\bi x};\i\omega R-1)
\end{equation}
obey the Klein-Gordon equation for the conformally-coupled
massless field over the ${\Bbb R}^1 \times H^3$ space.
Computing the Jacobian of the transformation from
${\bi q}$ to ${\bi q}'$ and using~(\ref{transf-phi}) it is easy to show that
the two-point function
\[ {\cal W}^{(0)} ({\bi x},{\bi y};\omega)=
\int_{S^{2}} d^{2}q\,
\phi_{\bi q}({\bi x};\i\omega R-1)
\phi_{\bi q}({\bi y};-\i\omega R-1) \]
is $SO(3,1)$-invariant:
\[ {\cal W}^{(0)}({\bi x}_{g},{\bi y}_{g};\omega)=
{\cal W}^{(0)}({\bi x},{\bi y};\omega). \]
Using~(\ref{hyperg-spin0})  we can write it the explicit form:
\begin{equation}\label{twop-0}
{\cal W}^{(0)}({\bi x},{\bi y};\omega)=
\frac{4\pi\sin \omega R\alpha(x,y)}{\omega R \sinh \alpha(x,y)}
\end{equation}
 where $\alpha(x,y)$ is the geodesic distance between the points:
$\cosh\alpha (x,y)=-R^{-2} x_\alpha y^\alpha$.
The above expression  coincides with the spherical function for
the spin-zero infinite-dimensional representations of the Lorentz
group obtained in~\cite{49}.

The equality
\begin{equation}\label{full}
\frac{1}{(2\pi)^3} \int \frac{d^3 x}{\kappa_{\bi x}}
\phi_{\bi q}({\bi x};\i\omega R-1)\phi_{{\bi q}'}({\bi x};-\i\omega' R-1)=
\frac{1}{4}\delta({\bi n}-{\bi n}')
\end{equation}
holds~\cite{Verd}.

\section{Spinor coherent states}

Let $\cal H$ be a little Lorentz group of the vector
$n^\alpha =(0,0,1,1)$. It is easy to show that its generators are
 $J_{10} +J_{13} ,\ J_{20} +J_{23}$ and $J_{12}$. Denote
  the three-dimensional rotation around the axis lying in the $xy$
plane as $g_{xy}$ and   parametrize these rotations by the three-vector
  $\bi q$ which is the result of action of $g_{xy}$ onto the standard
three-vector $(0,0,1)$. Expression for the
$g_{xy}({\bi q})$ in the $SL(2,{\Bbb C})$-form is~\cite{40}
\[ \left(
 \begin{array}{ll}
a & b \\ c & d
\end{array}
\right)=\frac{1}{\sqrt{2(1+q^3 )}}
\left(
\begin{array}{ll}
1+q^3 & q^1 -iq^2 \\
-q^1 -iq^2 & 1+q^3
\end{array}
 \right).\]
Denote  the boost along the third axis transforming the 
four-vector $(0,0,1,1)$ into the vector $(0,0,\omega,\omega)$
as $h_z =h_z (\omega)$; $SL(2,{\Bbb C})$-form of this transformation is
\[ h_z (\omega)=\left(
\begin{array}{ll}
e^{\omega/2} & 0 \\
0 & e^{-\omega/2}
\end{array}
\right).\]
Then the transformation $g_{xy}({\bi q})h_z (\omega)$ transforms
the vector $(0,0,1,1)$ into the vector~(\ref{lor5}) and then the mapping
\[ n^\alpha =(\omega{\bi q},\omega) \mapsto g_{xy}({\bi q})h_z (\omega)
\in SO(3,1)\]
composes the lifting from the $SO(3,1)/{\cal H}$ space to the Lorentz group.
From~(\ref{pi-s}), (\ref{lor2})
it follows that in the finite-dimensional representation $(j_+ ,j_-)$ the 
vector, being $\cal H$-invariant to within a phase multiplier, is
$|\psi_0 \rangle =1$. Then we can construct the CS system
\begin{equation}\label{lor8}
|{\bi q}\omega ;j_+ j_- \rangle =
T_{j_+ j_-} (g_{xy}({\bi q})q_z (\omega)) |\psi_0 \rangle.
\end{equation}
Using~(\ref{lor1}) it is easily seen that
\begin{equation}\label{lor6}
|{\bi q}\omega ;j_+ j_- \rangle =\left( \frac{n^4 +n^3}{2}\right)^{j_+ + j_-}
(1-z\rho_{\bi q})^{2j_+}
(1-\bar{z}\bar{\rho}_{\bi q})^{2j_-}
\end{equation}
where  $\rho_{\bi q}=\frac{q^{1}+iq^2}{1+q^3}$.
The transformation property of the constructed CS system is
\begin{equation}\label{4. 33}
T_{j_+ j_-}(g) |{\bi q}\omega ;j_+ j_- \rangle  \sim
\left( \frac{n^4_g}{n^4} \right)^{j_+ +j_-}
|{\bi q}'\omega ;j_+ j_- \rangle  \qquad {\bi q}'={\bi q}_{g^{-1}}
\end{equation}
where the equivalency relation $\sim$ is the equality to within a phase
multiplier. Indeed, the transformation
$T_{j_+ j_-}(g)$ transforms the vector
$|{\bi q}\omega ;j_+ j_- \rangle$ into the vector equivalent to
$|{\bi q}'\omega' ;j_+ j_- \rangle$. Since the vectors
$|{\bi q}\omega ;j_+ j_- \rangle$ depend on $\omega$ as $\omega^{j_+ +j_-}$
and $\omega=n^4$, then from here the equality~(\ref{4. 33}) follows.

In fact, the CS system~(\ref{lor6})  was implicitly constructed as
 basic spinors for the finite-dimensional representations of the Lorentz
group in the classical S.Weinberg's paper~\cite{48}.  In this paper the basic
spinors were determined by the relation equivalent to~(\ref{lor8}), and it
was shown that the vector  $|\psi_0 \rangle$ should be stable under the
little Lorentz group to within a phase multiplier. In~\cite{40} the
realization~(\ref{lor1}) of the Lorentz group irreducible representations was
used to the explicit construction of the vectors
$T_{j_+ j_-} (g_{xy}({\bi q})) |\psi_0 \rangle$.

Now we construct the functions
\begin{equation}\label{func-arbitspin}
f^{(s)}_{\bi q\omega}({\bi x})=\phi_{\bi q}({\bi x}; \i\omega R-s-1)
|{\bi q}\omega;s0\rangle .
\end{equation}

Then we can show~\cite{MPLA} that the functions
\[ \psi({\bi x},t)=\e^{\i\omega t}f^{(s)}_{\bi q\omega}({\bi x}) \]
obey the massless wave equations for arbitrary spin $s$. Using~(\ref{4. 33})
and~(\ref{transf-phi}) we obtain the transformation properties
\[ f^{(s)}_{{\bi q}\omega}({\bi x}_g)\sim
\left( \frac{n^4_g}{n^4} \right)^{\i\omega R-1}T_{s0}(g)
f^{(s)}_{{\bi q}'\omega}({\bi x})
\qquad {\bi q}'={\bi q}_{g^{-1}}.\]
Then the two-point function
\[ {\cal W}^{(s)}({\bi x},{\bi y};\omega)= \int_{S^{2}}
d^{2}q\, f_{\bi q\omega}^{(s)}({\bi x}) \otimes \left( f_{\bi
q\omega}^{(s)}({\bi y})\right)^{\dagger}\]
is $SO(3,1)$-invariant:
\[ {\cal W}^{(s)}({\bi x}_{g},{\bi y}_{g};\omega)=
T_{s0}(g){\cal W}^{(s)}({\bi x},{\bi y};\omega)T^{\dagger}_{s0}(g). \]

\section{Quantized fields}

Now we consider the quantized massless fields of arbitrary spin $s$ over the
 ${\Bbb R}^1 \times H^3$ space:
\begin{eqnarray}
\Psi^{(s)}(x)=\frac{2^{s-1/2}}{(2\pi)^{3/2}}
\int_{0}^{\infty}\omega d\omega
\int_{S^2}d^{2}q \nonumber\\
\times \left(
C_s (\omega)\e^{\i\omega x^0}f^{(s)}_{\bi q\omega}({\bi x})
a^{(+)}({\bi n};s)+
C_s (-\omega)\e^{-\i\omega x^0}f^{(s)}_{\bi q,-\omega}({\bi x})
a^{(-)\dagger}({\bi n};s)\right)\nonumber
\end{eqnarray}
where the four-vector $n^\alpha$ is given by~(\ref{lor5}), the
creation-annihilation operators obey the nonvanishing
(anti)commutation relations
\[ [a^{(+)} ({\bi n};s),a^{(+)\dagger} ({\bi n}';s)]_{\pm}=
[a^{(-)} ({\bi n};s),a^{(-)\dagger} ({\bi n}';s)]_{\pm}=
n^4 \delta^{3}({\bi n}-{\bi n}')\]
$C_0 (\omega)=1$ and
\[ C_s (\omega)=(1+(2\i \omega R)^{-1}) (1+3(2\i \omega R)^{-1}) \ldots
(1+s(\i \omega R)^{-1}) \]
at $s=1/2,3/2,\ldots$ and
\[ C_s (\omega)=(1+(\i \omega R)^{-1}) (1+2(\i \omega R)^{-1}) \ldots
(1+s(\i \omega R)^{-1}) \]
at $s=1,2\ldots$ Then the matrix elements of
corresponding propagators in the basis~(\ref{basis-j+j-}) are
\begin{equation}\label{prop-arbitspin}
 \fl D^{(s)}_{lm}(x,y)\equiv
[\Psi^{(s)}_{l}(x),\Psi^{(s)\dagger}_{m}(y)]_{\pm} = \frac{2^{2s-1}}{(2\pi)^3}
\int_{-\infty}^{\infty}\omega d\omega |C_s (\omega)|^2
e^{\i\omega(x^{0}-y^{0})}{\cal W}^{(s)}_{lm}({\bi x},{\bi y};\omega). 
\end{equation}
The above propagators are obviously
invariant under the time translations and spatial
$SO(3,1)$-transformations. In Appendix B we show that for an arbitrary
spin they obey the causality principle i.e. they are equal to zero at
 $(x^0 -y^0)^2 -R^2 (\alpha (x,y))^2 \not= 0$. In this attitude these
propagators are similar to those for massless fields of arbitrary spin
over the Anti-de Sitter space constructed in~\cite{Lesimple}. In Appendix C we show that at 
$R\rightarrow\infty$ these propagators pass onto the usual massless arbitrary spin propagators 
over Minkowski space constructed in~\cite{48}.

Using~(\ref{twop-0}) for the spin zero propagator we get~\cite{MPLA}
\begin{equation}\label{D0}
 D^{(0)}({\bi x},t_1 ;{\bi y},t_1 +t)=-
\frac{ \i\alpha}{2\pi\sinh\alpha}
\delta (t^2 -R^2 \alpha^2)\varepsilon (t).
\end{equation}
The above expression coincides with the difference of the
positive- and negative-frequency Wightmann functions, obtained previously by
the standard methods~\cite{33}, to within the multiple $1/4$.

Using the equalities~(\ref{hyperg-spin1/2})
we obtain the spin 1/2 two-point function
\begin{eqnarray}
{\cal W}^{(1/2)}({\bi x},{\bi y},\omega)=\i (C_{1/2}(\omega))^{-1}
(\kappa_{\bi x} +R^{-1}\bsigma{\bi x}) \nonumber \\
\times\left(\frac{1}{2R}-\i\omega +\sigma_i e_{(i)}^k
\frac{\partial}{\partial x^k}\right)
{\cal W}^{(0)}\left({\bi x},{\bi y},\omega-\frac{\i}{2R}\right). \nonumber
\end{eqnarray}
The corresponding propagator may be represented in the form
\begin{equation}\label{D1/2}
 D^{(1/2)}(x,y)=-2\i \e^{t/2R}
(\kappa_{\bi x}+R^{-1}\bsigma{\bi x})\left( \frac{\partial}{\partial x^0}
-\sigma_i e_{(i)}^k \frac{\partial}{\partial x^k} \right)
 D^{(0)}(x,y).
\end{equation}
At $R\rightarrow\infty$ the propagators~(\ref{D0}),(\ref{D1/2}) pass onto the 
usual massless spin 0 and 1/2 propagators 
over Minkowski space constructed in~\cite{48}.

\section{Discussion}

The common scheme of construction of quantum fields in curved space-time is the following~\cite{33}.
Let $I,J$ be the indices denoting the set of quantum numbers which distinguishes the states of
our particles; these indices  may run both discrete
and continuous sets of values. Define two sets of creation-annihilation operators 
with nonvanishing (anti)commutation relations
\[  [a_I ,a^\dagger_J ]_\pm = [b_I ,b^\dagger_J ]_\pm = \delta_{IJ} .\]
Introduce two sets $\varphi^{(\pm)}_I (x)$ of solutions of corresponding relativistic wave equations;
these solutions should be orthonormal with respect to the appropriate
scalar product:
\begin{eqnarray}\label{orth}
(\varphi^{(+)}_I ,\varphi^{(+)}_J)=(\varphi^{(-)}_I ,\varphi^{(-)}_J)=\delta_{IJ}  \\
(\varphi^{(+)}_I ,\varphi^{(-)}_J)=0 \label{orth+-}
\end{eqnarray}
and should be of positive (negative) frequency with respect to the
time-like Killing vector~$\xi$:
\begin{equation}\label{pos-neg}
{\cal L}_\xi \varphi^{(\pm)}_I =\pm \i\omega_I \varphi^{(\pm)}_I .
\end{equation}
Equality~(\ref{orth}) means the separation between
the particles with different quantum numbers, and the equalities~(\ref{orth+-}),(\ref{pos-neg})
mean the  separation between particles and antiparticles. 
Then we can define the quantized field as
\begin{equation}\label{field}
\Phi (x) =\sum\limits_I (\varphi^{(+)}_I (x) a_I + \varphi^{(-)}_I
(x) b_I^\dagger )
\end{equation}
and its propagator is equal to
\begin{equation}\label{prop}
D(x,y)\equiv [\Phi (x),\Phi^\dagger (y)]_\pm =\sum\limits_I
(\varphi^{(+)}_I (x) \varphi^{(+)\dagger}_I (y) - 
\varphi^{(-)}_I (x) \varphi^{(-)\dagger}_I (y) ).
\end{equation}
However, the area of applicability of such a scheme turns to be limited. Usually it is impossible to perform 
the summation/integration at the r.h.s of~(\ref{prop}) in the closed form,
especially  for the fields of nonzero spin.
In the cases when it is possible, the resulting propagator may violate the causality principle,
as in the case of massive spin 1/2, 1 and 2 fields over de Sitter space~\cite{Spindel}. From the other
 hand, for the spin zero massless minimally coupled field over the same space it is necessary to introduce
the noninvariant vacuum~\cite{Allen} or states with negative norm~\cite{Gaz}.

Another serious difficulty comes from the notion of particles.
Generaly speaking, it have not the invariant (independent from an observes) sense.
 Even in the cases
when the unambiguous separation between the positive- and negative-frequency modes
is possible (e.g. in the stationary space-times), the correct definition of particles 
may be still unavailable, as its takes place inside the 
Schwarzschild black hole~\cite{Schutz}. Even in the space ${\Bbb R}^1 \times
H^3$ whose metrics is stationary and regular everywhere, the original
pathology appears: the set of functions which obey~(\ref{orth}) at finite $R$,
does not obey this equality at $R\rightarrow\infty$ due to the multiplier 
$1/4$ at the r.h.s. of~(\ref{full}). Then the particles which are defined in the ${\Bbb R}^1 \times
H^3$ space due to~(\ref{orth}),(\ref{orth+-}) and~(\ref{pos-neg}), do not coincide with the usual particles 
over Minkowski space at $R\rightarrow\infty$, in spite of the fact that the plane waves~(\ref{psi-0})
in this limit pass into the usual plane waves over Minkowski space.

Then, constructing the QFT in curved space there is no reasons to start from the definition of particles 
given by~(\ref{orth}),(\ref{orth+-}) and~(\ref{pos-neg}). However, in some important cases such
as de Sitter and ${\Bbb R}^1 \times H^3$ symmetric
spaces, we can use another generalization of Minkowskian notion of
particles via appropriate generalization of the Poincar\'e invariance. Namely, 
we can construct the functions $\varphi^{(\pm)}_I$ as the "plane waves" which reflect the symmetry
properties of the space and generalize the usual plane waves over Minkowski space. Such an
approach is developed in~\cite{Bros,dS-JPA} for the massive spin 0, 1/2 and 1 fields over de Sitter 
space and in~\cite{MPLA} and in the present paper for the massless arbitrary spin fields over
the ${\Bbb R}^1 \times H^3$ space. In~\cite{dS-JPA,MPLA} and in the present paper we show that
the CS method is a natural mathematical tool for constructing these "plane waves" and studying their
properties.

In this approach the physical sensibility of the theory is based on the conditions imposed
on the two-point functions
\[ D^+ (x,y)=\langle 0|\Phi(x) \Phi^\dagger (y)|0\rangle = 
\sum\limits_I \varphi^{(+)}_I (x) \varphi^{(+)\dagger}_I (y) \]
rather on the functions $\varphi^{(\pm)}_I$ itself, where $\Phi(x)$ is still given by~(\ref{field}) and
$a_I |0\rangle =b_I |0\rangle =0$. 
In the case of the ${\Bbb R}^1 \times H^3$ space the function $D^+ (x,y)$ is  the positive
frequency part of the corresponding propagator.

Like~(\ref{orth}),(\ref{orth+-}) and~(\ref{pos-neg}), the mentioned 
conditions   generalize the corresponding Minkowski space conditions:

1) Invariance:
\[ D^+ (x_g ,y_g)=U(g)D^+ (x,y) \overline{U}(g) \]
where $g\in {\cal G}$ is an arbitrary element of the symmetry group $\cal G$ of our symmetric
space $X$; $X\ni x\mapsto x_g \in X$ is the action of $\cal G$ over this space, and $U(g)$ is
some finite-dimensional representation of $\cal G$.

2) Causality:
$D^+ (x,y) = D^+ (y,x)$ if the points $x,y$ are
spacelike-separated (massive case) or not lightlike-separated (massless case).

3) Positiveness:
\[ \int_X  \int_X \d\sigma (x) \d\sigma (y) \, f(x) D^+ (x,y) f^* (y) \geq 0 \]
where $\d\sigma(x)$ is the $\cal G$-invariant measure on $X$ and $f(x)$ is an
arbitrary $C^\infty$-function with compact support in $X$.

4) Analyticity: $D^+ (x,y)$ is a boundary value of some analytic function defined in the certain
domain of the appropriate complexification of $X$.

Then the Hilbert space structure of field theory may be obtained via the reconstruction
theorem~\cite{Streat}.
In~\cite{Bros,dS-JPA} it was shown that the above properties hold for the massive spin 0, 1/2 and 1
fields over de Sitter space. Within the approach advocated in~\cite{dS-JPA,MPLA} and in the 
present paper, the properties 1) and 3) immediately follow from the construction of $D^+ (x,y)$.
In Appendix B  we show that the properties 2) and 4) also hold for the massless arbitrary spin 
fields over the ${\Bbb R}^1 \times H^3$ space. 
Thus, the theory of these fields constructed above
obeys all the requirements  to the meaningful free QFT in curved space.

\appendix
\section{Two-point functions}

The polynomials
\begin{equation}\label{basis-j+j-}
|m\rangle =(C_{2s}^{m-1})^{1/2}z^{m-1} \qquad m=1,\ldots,2s+1
\end{equation}
compose the orthonormal basis in the spin $s$ representation of the $SO(3)$
group. We can consider them as the columns:
\[ |1\rangle =\left[
\begin{array}{l}
1 \\ 0 \\ 0 \\ \vdots \\ 0
\end{array}
\right] \qquad
|2\rangle=\left[
\begin{array}{l}
0 \\ 1 \\ 0 \\ \vdots \\ 0
\end{array}
\right] \ldots \]
We can  obtain the  arbitrary spin two-point function for the points
\[ x^\alpha_\circ =(0,0,R\sinh\alpha ,R\cosh\alpha) \qquad
y^\alpha_\circ =({\bi 0},R).\]
Then passing to the spherical coordinates over the 2-sphere and using
the formula~\cite{59}
\[ \mathop{_2 F_1} (a,b;c;z)=
\frac{2^{1-c}\Gamma(c)}{\Gamma(b)\Gamma(c-b)}
\times \int_{0}^\pi d\varphi
\, \frac{(\sin\varphi)^{2b-1}(1+\cos\varphi)^{c-2b}}{\left(
1-\frac{z}{2}+\frac{z}{2}\cos\varphi\right)^a}\]
we obtain
\begin{equation}\label{twop-h3-arbitspin}
\fl {\cal W}^{(s)}_{lm}({\bi x}_\circ ,{\bi y}_\circ ;\omega)=
\frac{4\pi\omega^{2s}}{2s+1} \e^{(-\i\omega R+s+1)\alpha}\delta_{lm}
\mathop{_2 F_1}(-\i\omega R+s+1,l;2s+2;1-\e^{2\alpha}).
\end{equation}
In~\cite{coher1} for the scalar product of two spin zero CS the expression
\[ {\cal W}^{(0)}({\bi x},{\bi y};\omega)=
4\pi\mathop{_2 F_1}\left(-\frac{\i\omega R-1}{2},
\frac{\i\omega R+1}{2};\frac{3}{2};
-\sinh^2 \alpha(x,y)\right)\]
was obtained. It is a particular case of~(\ref{twop-h3-arbitspin})
since the equality~\cite{59}
\[ \mathop{_2 F_1}(a,b;2b;z)=
(1-z)^{-a/2}\mathop{_2 F_1}\left( \frac{a}{2},b-\frac{a}{2};b+\frac{1}{2};
\frac{z^2}{4(z-1)}\right) \]
holds. At $s=0$
\begin{equation}\label{hyperg-spin0}
\mathop{_2 F_1} (-\i\omega R+1,1;2;1-\e^{2\alpha})=
\frac{\e^{2\i\omega R\alpha}-1}{\i\omega R(\e^{2\alpha}-1)}.
\end{equation}
At $s=1/2$
\begin{equation}
\label{hyperg-spin1/2}
\eqalign{
\fl\mathop{_2 F_1} \left( -\i\omega R+\frac{3}{2},1;3;1-\e^{2\alpha}\right)=
-2\frac{\e^{2\i\omega R\alpha}\e^\alpha+
\left(\i\omega R+\frac{1}{2}\right) (1-\e^{2\alpha})-1}
{\left(\omega^2 R^2 +\frac{1}{4}\right)(1-\e^{2\alpha})^2} \\
\fl\mathop{_2 F_1} \left( -\i\omega R+\frac{3}{2},2;3;1-\e^{2\alpha}\right)=
2\frac{\e^{2\i\omega R\alpha}\e^\alpha \left(\i\omega R-\frac{1}{2}\right)
-\left( \i\omega R+\frac{1}{2}\right)\e^{-\alpha}+1}
{\left(\omega^2 R^2 +\frac{1}{4}\right)(1-\e^{2\alpha})^2} .
}
\end{equation}

\section{Causality}

We denote
\[ K^{(s)}_l (\alpha,\omega)=\omega |C_s (\omega)|^2
{\cal W}^{(s)}_{ll}({\bi x}_\circ ,{\bi y}_\circ ;\omega).\]
Twice applying  the expression~\cite{59}
\[ \fl\frac{(-1)^n (a)_n (c-b)_n }{(c)_n}(1-z)^{a-1}
\mathop{_2 F_1} (a+n,b;c+n;z)=
\frac{d^n}{dz^n}\left((1-z)^{a+n-1}
\mathop{_2 F_1} (a,b;c;z) \right) \]
where $(a)_n =\Gamma (a+n)/\Gamma(a)$ and using the hypergeometric equation
we obtain
\begin{equation}\label{a1b1c2}
\eqalign{
\frac{a(c-a)b(c-b)}{c(c+1)}\mathop{_2 F_1} (a+1,b+1;c+2;z) \\
\lo= \left(\frac{d^2}{dz^2}+(c-a-b-1)\frac{d}{dz}\right)
\mathop{_2 F_1} (a,b;c;z).
}
\end{equation}
Then
\begin{equation}\label{K}
K_l^{(s)}(\alpha,\omega)=p_{s,l}(\alpha,\omega)\e^{\i\omega R\alpha}+
q_{s,l}(\alpha,\omega)\e^{-\i\omega R\alpha}
\end{equation}
where $p_{s,l}(\alpha,\omega)$ and $q_{s,l}(\alpha,\omega)$ are the
polynomials on $\omega$. Indeed, the validity of~(\ref{K}) at
$s=0$ and $s=1/2$ follows from~(\ref{hyperg-spin0})
and (\ref{hyperg-spin1/2}) immediately. Let~(\ref{K}) be valid for
$K^{(s)}_l ,\ l=1,\ldots,2s+1.$ Then its validity for
$K^{(s+1)}_l$ at $l=2,\ldots,2s+2$ follows from~(\ref{a1b1c2}).
From the other hand, the validity of~(\ref{K}) for
$K^{(s+1)}_{1}$ and $K^{(s+1)}_{2s+3}$ follows from its validity for
$K^{(s+1)}_{2}$ and $K^{(s+1)}_{2s+2}$ using the expressions~\cite{59}
\begin{eqnarray}
(a)_n z^{a-1}\mathop{_2 F_1} (a+n,b;c;z)=\frac{d^n}{dz^n}
\left(z^{a+n-1}\mathop{_2 F_1} (a,b;c;z) \right) \nonumber \\
(c-a)_n z^{c-a-1}(1-z)^{a+b-c-n} \mathop{_2 F_1} (a-n,b;c;z) 
\nonumber \\
\lo= \frac{d^n}{dz^n}\left(z^{c-a+n-1}(1-z)^{a+b-c}
\mathop{_2 F_1} (a,b;c;z) \right) .\nonumber
\end{eqnarray}
Thus~(\ref{K}) is proved for the arbitrary $s$ and $l$. Then for the
propagator we obtain
\begin{equation}\label{caus}
\eqalign{
D^{(s)}_{lm}(t_1 ,{\bi x}_\circ ;t_1 +t, {\bi y}_\circ) =
\delta_{lm}\frac{2^{2s+2}\pi^2}{2s+1}  \\
\fl\times \left[
p_{s,l}\left(\alpha,-\i\frac{\partial}{\partial (t+R\alpha)} \right)
\delta(t+R\alpha)+
q_{s,l}\left(\alpha,-\i\frac{\partial}{\partial (t-R\alpha)} \right)
\delta(t-R\alpha)\right] .
}
\end{equation}
where we consider the differential operators in the arguments of
$p_{s,l}$ and $q_{s,l}$ do not acting onto the first arguments of these
functions. Then the propagator is not equal to zero only at
$|t|=R\alpha ({\bi x}_\circ ,{\bi y}_\circ)$; then its causality for the
arbitrary ${\bi x}$ and $\bi y$ follows from its Lorentz-invariance.

From~(\ref{hyperg-spin0}), (\ref{hyperg-spin1/2}) and~(\ref{a1b1c2})
it follows that the functions $p_{s,l}(\alpha,\omega)$ and $q_{s,l}(\alpha,\omega)$ have not 
singularities at $\alpha\not= 0$. Consider the domain
\begin{equation}\label{anal}
\{(x,y)| \Im (x^0 -y^0)>0, \ {\bi x},{\bi y}\in {\Bbb R}^3,\ \alpha(x,y)\not=0\}.
\end{equation}
Since the positive-frequency part of the function $\delta(x)$ is a boundary value of the function
$\i\pi^{-1}x^{-1}$ at $\Im x\rightarrow +0$, then the two-point functions $D^{+(s)}(x,y)$ are analytic
in the domain~(\ref{anal}),  and they may be obtained from the r.h.s. of~(\ref{caus})  replacing
$\delta (t\pm R\alpha)$ to $\i\pi^{-1}(t\pm R\alpha)^{-1}$.

\section{Minkowski space limit}

Since the functions~(\ref{func-arbitspin}) at $R\rightarrow\infty$
pass onto the usual plane waves over Minkowski space constructed 
in~\cite{48,40}, then we can expect that the analogous  passage
takes place for the propagators too. However, since the corresponding integrals 
makes sense in the terms of generalized functions, then this
 limiting passage demands additional justification (cf. discussion of equation~(\ref{full})
in Section 6). Define
\[ \tilde{D}\vphantom{D}^{(s){\rm flat}}(x-y)=
\lim_{R\rightarrow\infty}D^{(s)}(x,y) \]
where $D^{(s)}(x,y)$ is given by~(\ref{prop-arbitspin}). The propagators
for the arbitrary spin massless fields over  Minkowski space are~\cite{48}
\[ D^{(s)\rm flat}(x)=\frac{2^{2s-1}}{(2\pi)^3}\int_{n^2 =0}\frac{d^3 n}{n^0}
e^{\i n\cdot x} |{\bi q}\omega ;s0\rangle\langle {\bi q}\omega ;s0|.\]
Now we prove that $\tilde{D}\vphantom{D}^{(s){\rm flat}}(x)=D^{(s)\rm flat}(x)$. At $s=0$
and  $s=1/2$ it was proved in Section 5. Then we only should prove that the matrix 
elements of propagators $\tilde{D}\vphantom{D}^{(s){\rm flat}}(x)$ and
$\tilde{D}\vphantom{D}^{(s+1){\rm flat}}(x)$ are connected with each other 
by the same recurrence relations as the matrix elements of 
$D^{(s)\rm flat}(x)$ and $D^{(s+1)\rm flat}(x)$ are. To this end we put
$x^1 =x^2 =0$. Then in the basis~(\ref{basis-j+j-}) we obtain
\[ D^{(s)\rm flat}_{lm}(x)=\frac{C_{2s}^{m-1}}{2(2\pi)^3} \delta_{lm}
 \int_{n^2 =0}\frac{d^3 n}{n^0} e^{\i n\cdot x}(n^0
-n^3)^{2s+1-m}(n^0 +n^3)^{m-1}. \]
From the above expression the recurrence relations
\begin{equation}\label{recurrent}
\eqalign{
(\partial_0 -\partial_3 )^2 D^{(s)\rm flat}_{mm}(x)
=-(2s+2-m)(2s+3-m) D^{(s+1)\rm flat}_{mm}(x)\\
(\partial_0^2 -\partial_3^2 )D^{(s)\rm flat}_{2s+1 \ 2s+1}(x)
=-\frac{1}{2s+2}D^{(s+1)\rm flat}_{2s+2 \ 2s+2}(x) \\
(\partial_0 +\partial_3 )^2 D^{(s)\rm flat}_{2s+1 \ 2s+1}(x)=-
D^{(s+1)\rm flat}_{2s+3 \ 2s+3}(x)
}
\end{equation}
follow immediately. From the other hand, in Appendix B we show 
that the integrand in the r.h.s. of equation~(\ref{prop-arbitspin}) 
is a polynomial on $\omega$. Then we can bring the limiting passage 
inside the integral on $\omega$. Then using~(\ref{twop-h3-arbitspin}) 
we obtain
\[ \tilde{D}\vphantom{D}^{(s)\rm flat}_{lm}(x)
=\frac{2^{2s-1}}{2\pi^2 (2s+1)} \delta_{lm}
\int_{-\infty}^\infty d\omega \,
\omega^{2s+1}e^{\i\omega (x^0 -x^3)}\mathop{_1 F_1}(m,2s+2,2\i\omega x^3).\]
Then the validity of recurrence relations~(\ref{recurrent}) for the 
propagators $\tilde{D}\vphantom{D}^{(s)\rm flat}(x)$ follows from the 
formulae~\cite{59}
\[\eqalign{
\frac{d^n}{dx^n}\mathop{_1 F_1}(a,c;x)=
\frac{a(a+1)\ldots (a+n-1)}{c(c+1)\ldots (c+n-1)}\mathop{_1 F_1} (a+n,c+n;x)
\\
\fl \frac{d^n}{dx^n}\left( e^{-x}\mathop{_1 F_1}(a,c;x)\right)= 
\frac{(c-a)(c-a+1)\ldots (c-a+n-1)}{c(c+1)\ldots (c+n-1)}e^{-x}\mathop{_1
F_1} (a,c+n;x)
} \]
and the formula
\[\frac{1}{a}\mathop{_1 F_1}(a,a+1;x)-\frac{1}{a+1}\mathop{_1 F_1}(a+1,a+2;x)
=\frac{1}{a(a+1)}\mathop{_1 F_1}(a,a+2;x)\]
which may be easily derived from the relations between the adjacent 
confluent hypergeometric functions~\cite{59}.

\end{document}